\documentclass[pra,twocolumn,tightenlines,footinbib,floatfix,
superscriptaddress,showpacs,amsmath,amssymb]{revtex4}

\usepackage{amssymb}
\usepackage{amsmath}
\usepackage{colordvi}
\usepackage{amsthm}
\usepackage{epsfig}
\def\avg#1{\langle#1\rangle}

\def\be{\begin{equation}}
\def\ee{\end{equation}}
\def\bea{\begin{eqnarray}}
\def\eea{\end{eqnarray}}

\def\nn{\nonumber}

\begin{document}
\title{Vortex structures of rotating spin-orbit coupled Bose-Einstein
condensates}
\author{Xiang-Fa Zhou}
\affiliation{\textit{Department of Physics, University of
California, San Diego, CA 92093}}
\affiliation{\textit{Key Laboratory of Quantum Information,
University of Science and Technology of China, CAS, Hefei, Anhui
230026, People's Republic of China}}
\author{Jing Zhou}
\affiliation{\textit{Key Laboratory of Quantum Information,
University of Science and Technology of China, CAS, Hefei, Anhui
230026, People's Republic of China}}
\author{Congjun Wu}
\affiliation{\textit{Department of Physics, University of
California, San Diego, CA 92093}}

\begin{abstract}
We consider the quasi-$2$D two-component Bose-Einstein condensates with
Rashba spin-orbit (SO) coupling in a rotating trap.
The rotation angular velocity couples to the mechanical angular momentum
which contains a non-canonical part arising from SO coupling.
The effects of an external Zeeman term favoring spin polarization
along the radial direction is also considered, which has the same form as
the non-canonical part of the mechanical angular momentum.
The rotating condensate exhibits a variety of rich structures as varying the
strengths of the trapping potential and interaction.
With a strong trapping potential, the condensate exhibits a
half-quantum vortex-lattice configuration.
Such a configuration is driven to the normal one by introducing
the external radial Zeeman field.
In the case of a weak trap potential, the condensate exhibits a
multi-domain pattern of plane-wave states under the external radial
Zeeman field.
\end{abstract}
\maketitle

\section{Introduction}
\label{sect:intro}

Spin-orbit (SO) coupling plays an important role in various aspects in
condensed matter systems including spintronics \cite{spintronics} and
topological insulators \cite{tpinsulator,qi2010}.
However, SO effects in bosonic systems has not been attracted much
attention until recently.
For example, $^4\mbox{He}$ atoms are spinless and ultracold spinful
bosons are too heavy to exhibit relativistic SO coupling. This situation is
significantly changed by the recent experimental progress in both
semiconductor exciton systems and cold atom systems with synthetic
gauge fields. Excitons are composite bosons of electrons and holes.
Their effective masses are light enough to exhibit relativistic SO
coupling.
Exotic SO coupled condensates with stripe and skyrmion types spin texture
configurations was theoretically predicted by Wu and Mondragon-Shem
\cite{wu2008}.
Excitingly, spin textures have been observed in the SO coupled
exciton condensates by High {\it et al} \cite{butov}.
On the other hand, many theoretical schemes have been proposed in
ultracold atomic systems to create artificial non-Abelian gauge
fields by using laser-atom interactions
\cite{scheme0,scheme1,scheme2,scheme3,scheme4,
scheme5,scheme6,scheme7, scheme8,scheme9,scheme10,scheme11},
which generate effective SO coupling without special relativity.

It has been shown that bosons with SO coupling support exotic ground
states beyond the ``no-node" theorem \cite{wu1,nontrival-top,yao2008,
wu2009}.
This theorem  states that the ground state wavefunctions of bosons
under very general conditions are positive definite, which
is essentially a direct result
of the Perron-Frobenius theorem of matrix analysis
\cite{bapat1997}.
However, the linear coupling to momentum in the SO coupling invalidates the
proof of the ``no-node'' theorem.
For example, spontaneous time-reveal symmetry breaking states
exhibiting spin-density wave ordering \cite{wu2008,wu1,zhai,ho,yip,zhang}
and spontaneous half-quantum vortex configuration \cite{wu1,wu2008}
have been studied.
Both of them exhibit either nodal or complex-valued
condensate wavefunctions, and thus are beyond the ``no-node'' theorem.
Especially, the realization of SO coupled Bose-Einstein condensations
(BEC) of $^{87}$Rb \cite{exp1,exp2}
provides a valuable opportunity to investigate this type of exotic
physics, experimentally.
Another way to bypass ``no-node'' theorem is to employ the meta-stable
excited states, in which ``no-node'' theorem does not apply either.
For example,  cold alkali bosons have been pumped into
the high orbitals in optical lattices \cite{mueller2007,wirth2010}.
It was shown that interactions among $p$-orbital bosons
obey an ``orbital Hund's rule'', which generates a class of
orbital superfluid states with complex-valued wave functions
breaking TR symmetry spontaneously \cite{nonode1,nonode2,
nonode3,nonode4,nonode5}.

On the other hand, vortex properties in rotating BECs are a
characteristic topological feature of superfluidity including $^4$He
and ultra-cold bosons, which have been studied extensively both
experimentally and theoretically \cite{fetter}. For spinor BECs and
spinful Cooper pairing superfluidity (e.g. superfluid $^3$He A and
B-phases), exotic spin textures and fractional quantized vortices
can form under rotation \cite{spinor}. However, to our knowledge,
the vortex properties of rotation SO coupled BECs have not been
thoroughly investigated before.

In this article, we investigate the rotating SO coupled condensate in
a quasi-2D harmonic trap with the angular velocity along the $z$-axis.
The angular velocity couples to the mechanical angular momentum
whose non-canonical part behaves like a Zeeman term polarizing
spin in the radial direction.
We also consider the effect from an external Zeeman term with
the same form.
The single particle ground states in the absence of interaction
can have non-zero vortex numbers, which differ by one
in the spin-up and down components as a result of SO coupling.
With many-body interactions, the rotating condensate exhibit
a variety of configurations depending on the strengths of the
trapping potential and interaction.
If the trapping potential is strong and interaction is relatively
weak, a half-quantum vortex lattice is formed under rotation.
Its spin configuration is a lattice of skyrmions.
The condensate of the spin up component breaks into disconnected
density peaks, which overlap the vortex cores of the spin-down condensate.
The presence of the external Zeeman field drives the system from a
half quantum vortex lattice state to a normal quantum vortex lattice
state.
In the case of a weak trap potential, the condensate favors a plane-wave
state or a two-plane-wave state with twist phase profiles under rotation.
With the external Zeeman field, the condensate develops multi-domain
configuration of plane-wave states.
The configuration of wavevectors can be clockwise or counter-clockwise
depending on the direction of the field.

The rest part of the paper is organized as follows.
The model Hamiltonian of the rotating Rashba coupled BEC is introduced
in Sect. \ref{sect:hamltn}.
The solution of the single particle wavefunction is
presented in Sect. \ref{sect:single}.
The rich structures of the vortex configurations with spin textures
are given in Sect. \ref{sect:vort}.
onclusions are given in  Sect. \ref{sect:conclusion}.

\section{The Model Hamiltonian}
\label{sect:hamltn}

We consider the quasi-2D two-component BECs with Rashba SO coupling in
the $xy$-plane subject to a rotation angular velocity $\Omega_z$ along
the $z$-direction.
The free part of the Hamiltonian of Rashba SO coupling under rotation
is defined through the standard minimal coupling as
\bea
\label{Hamiltonian}
H_0 &=& \int d^3 \vec{r} \psi_{\mu}^{\dag}(\vec{r})
\big [ \frac{1}{2 M} (-i \hbar \vec{\nabla} + M \lambda
\hat z \times \vec \sigma -
\vec{A})^2 -\mu
\label{eq:h0}
\nn \\
&+&V_{ext}(\vec{r})- \frac{1}{2}M\Omega_z^2 (x^2+y^2) \big]_{\mu \nu}
\psi_{\nu}(\vec{r}),
\eea
where $\vec \sigma = \sigma_x \hat x + \sigma_y \hat y +\sigma_z \hat z$ with $\sigma_{x,y,z}$ the usual Pauli matrices;
$\lambda$ is the Rashba SO coupling strength with the unit of velocity;
$\mu,\nu$ take values of $\uparrow,\downarrow$ as pseudospin
indices; $\vec{A}=(-M\Omega_z y, M \Omega_z x,0)$
is the vector potential from  Coriolis force;
$V_{ext} (\vec{r}) = \frac{1}{2} M \omega_T (x^2+y^2)$ is the
external harmonic trapping potential;
the last term in Eq. \ref{eq:h0} is the centrifugal force.
The interaction part $H_{int}$ is defined as
\bea
H_{int}&=& \frac{g_{\mu \nu}}{2}\int d^3 \vec{r}
\psi_{\mu}^{\dag}(\vec{r})
\psi_{\nu}^{\dag}(\vec{r})
\psi_{\nu}(\vec{r}) \psi_{\mu}(\vec{r}).
\eea
We assume the equal
intra-component interactions as $g_{\uparrow\uparrow} =g_{\downarrow
\downarrow} = g$, and inter-component interaction
$g_{\uparrow\downarrow} = gc$ with $c$ a constant coefficient.

Due to the presence of SO coupling, $\Omega_z$
couples to the mechanical angular momentum $L^{mech}$
rather than the canonical one $L_z$.
We extract this coupling from Eq. \ref{eq:h0} as
\bea
H_{rot}= - \Omega_z \int d^3 \vec{r}
\psi_{\mu}^{\dag}(\vec{r}) \Big[ L^{mech} \Big ]_{\mu\nu} \psi_{\nu}(\vec{r}),
\eea
where
\bea
L^{mech}=L_z+M\lambda (x \sigma_x +y\sigma_y).
\eea
Therefore, rotation in the presence of SO coupling induces an
effective magnetic field distribution $\vec{B}_{R}(\vec r)
= \Omega_z M \lambda (x, y,0)$ in the $xy$-plane.
As we will see below, this non-canonical
part in $L^{mech}$ plays a crucial role during the understanding the
single-particle ground state properties.

For the later convenience, we also introduce an external spatially dependent
Zeeman term as
\bea
H_B&=& - \int d^3r \psi^\dagger_\mu(\vec r) ( B_{ex,x} \sigma_x + B_{ex,y}
\sigma_y)_{\mu\nu} \psi_\nu (\vec r), \ \ \
\label{eq:zeeman}
\eea
where $\vec{B}_{ex}(\vec r)=(B_0x, B_0y,0)$ varies linearly in the $xy$-plane.
Such a term can tune the strength of the non-canonical part of the
mechanical momentum, which renders the model
adjustable in a wider range of the parameter space.

Many efforts have been made to implement the above Hamiltonian in
ultra-cold atomic gases.
Several schemes have been proposed to generate Rashba SO coupling
\cite{scheme0,scheme10,scheme11} with tunable SO coupling strength.
In particular, proposals in Ref. [\onlinecite{scheme10,scheme11}]
have the advantage to overcome the drawback of the spontaneous emission
in the tripod scheme. The spatially dependent Zeeman term $H_B$
can be generated through coupling two spin components
using two standing waves in the $x$ and $y$ directions with a
phase difference of $\pi/2$.
The resulting Rabi coupling is written as
\bea
-\Omega \big[ \sin(k_L x) +i \sin(k_L y) \big] \psi^\dagger_\downarrow(\vec r)
\psi_\uparrow(\vec r) + h.c.
\eea
In the region of $x, y\ll 2\pi/k_L$,
it reduces to the desired form of Eq. \ref{eq:zeeman} with
$B_0=\Omega k_L$.

\section{The single particle spectra}
\label{sect:single}
We start with the non-interacting Hamiltonian $H_0+H_B$, to gain
some intuition.
The confining trap is characterized by the length scale
$l=\sqrt{\hbar/M\omega}$.
We define another length scale $l_{so} =\hbar/(M\lambda)$ from SO
coupling.
The ratio between them $\alpha=l/l_{so}$ is dimensionless parameter
to describe the strength of SO coupling.
For the typical setup used in the NIST group \cite{exp2}, $\alpha \sim 10$.
Below we vary the values of $\alpha$ from $0 \sim 10$.
Experimentally, the regime of small $\alpha$ can be reached by using a
deeper trap potential.

If without the confining potential and rotation,
the single-particle eigenstates is of the form
\bea
\psi_{\pm,\vec{k}} =
e^{i\vec{k}\cdot\vec{r}} | \pm, \vec{k}\rangle,
\eea
where $|\pm, \vec{k} \rangle = \frac{1}{\sqrt{2}}
\left  ( 1, \mp e^{i \theta_{\vec{k}}} \right)^T$,
and $\theta_{\vec{k}}$ is the azimuthal angle of $\vec{k}$.
Since the condensate is uniform along the $\hat z$ direction,
we always have $k_z=0$ for the ground state.
The corresponding dispersion relations come into two branches
$\epsilon_{\pm}=\hbar^2(k^2\pm 2 k_0 k)/(2M)$ with
$k_0=1/l_{so}$.
Therefore, the single particle ground states are infinitely degenerate
along a ring in momentum space with radius $k_0$.

The external harmonic potential has an important effect which lifts
the degeneracy along the Rashba ring as pointed out in Ref. \cite{wu2008}.
In the momentum representation, the harmonic potential becomes
$ \frac{1}{2} M \omega^2 (i\hbar \nabla_{\vec{k}}-\vec{A'})^2 $
in the lower branch and couples different plane wave states around
the Rashba ring, where
$\vec A'(\vec k)=i \langle \psi_{-,\vec{k}} |\nabla_{\vec{k}}
|\psi_{-,\vec{k}} \rangle$ corresponding to a $\pi$-flux at the origin.
Therefore the motion along the Rashba ring is quantized
and maintains time-reversal (TR) invariance.
The single particle spectra exhibit the fermion-type Kramer degeneracy
with $T^2=-1$.
The lowest single particle eigenstates carry $j_z=\pm\frac{1}{2}$.
As shown in Ref. \cite{wu2008}, the angular quantization
gives rise to the dispersion on $j_z$ as
\bea
\frac{1}{\alpha^2} |j_z|^2 \hbar \omega_T.
\eea
On the other hand, the radial quantization is the same as in the
ordinary harmonic trap, which is at the order of $\hbar \omega_T$
\cite{wu2008}.
In the strong SO coupling limit, {\it i.e.}, $\alpha\gg 1$, the
dispersion over $j_z$ is nearly flat.
Thus the radial quantum number can be viewed as band index,
and the quantum number $j_z$ marks each state in the band.

\begin{figure}
\includegraphics[width=0.8\linewidth]{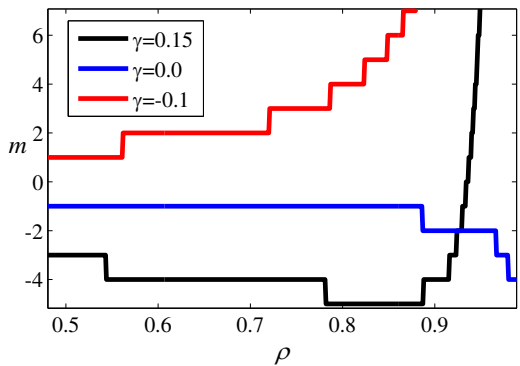}
\caption{The canonical angular momenta $m$ of the single
particle ground states described in Eq. \ref{eq:wavefunc}
{\it v.s.} $\rho$ for $\gamma=-0.1$, $0.0$, and $0.15$,
respectively.}
\label{fig:single_particle}
\end{figure}

To be more precise, we define two independent annihilation operators
as $\hat{a}_d = \frac{1}{2}( \bar{z} + 2 \partial_z )$ and
$\hat{a}_g = \frac{1}{2}( z + 2 \partial_{\bar{z}} )$ where
$z=(x\pm iy)/l$ and $\bar z$ is the complex conjugate of
$z$ \cite{shem,fetter}.
The single-particle Hamiltonian can be rewritten in the unit
$\hbar \omega$ as
\bea
H_0&+&H_B= (1-\rho)\hat{N}_d + (1+\rho) \hat{N}_g  + 1 \nn \\
&+& \alpha \left \{ [(1-\kappa)\hat{a}_d-(1+\kappa)
\hat{a}_g^{\dag}]\sigma^+ + h.c. \right \},
\label{eq:non-int}
\eea
where
\bea
\rho=\Omega_z/\omega, \ \hat{N}_d=\hat{a}_d^{\dag} \hat{a}_d, \
\hat{N}_g=\hat{a}_g^{\dag} \hat{a}_g, \ \sigma^+=
\frac{1}{2}(\sigma_x+i\sigma_y), \nn
\eea
and $\kappa=\gamma+\rho$ with $\gamma = B_0/(M \omega \lambda)$.
The corresponding canonical angular momentum reads $L_z=\hbar l_z=\hbar
(\hat{N}_d-\hat{N}_g)$.
The $\kappa$-term represents the combined effect from the non-canonical
part of $H_{rot}$ and the Zeeman term $H_B$.

\begin{figure}
\includegraphics[width=0.8\linewidth]{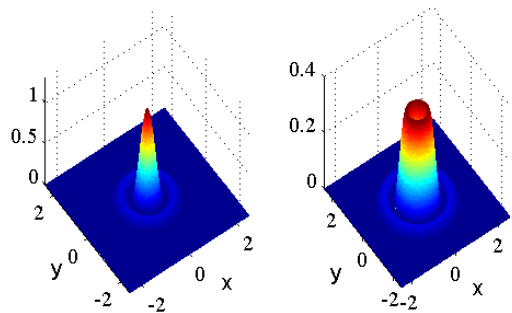}
\caption{Density profiles of spin-up and down components for the single-particle ground state
$\phi_{j_z = \frac{1}{2}}$ with the parameter values of $ \alpha = 4 $,
and $\rho=\gamma=0$.
Only the spin-down component carries a vortex.
The corresponding density profile for $\phi_{j_z = -\frac{1}{2}}$ is obtained
by interchanging the spin indices.}
\label{fig:single-particle-density}
\end{figure}

We diagonalize  Eq. \ref{eq:non-int} to obtain the single
particle spectra, and
present the solutions in the coordinate representation, in
which the ground state wavefunction reads as
\bea
e^{i m \phi } \left (
\begin{array}{c}
f(r) \\
g(r) e^{i \phi}
\end{array}
\right ).
\label{eq:wavefunc}
\eea
The total canonical angular momentum $j_z=l_z+\frac{1}{2}\sigma_z=m+\frac{1}{2}$
remains a conserved quantity, thus the canonical orbital angular
momenta in the two spin components differ by one due to SO coupling.
Fig. \ref{fig:single_particle} shows $m$ as a function of the rotational
angular velocity $\rho$ for different external magnetic field $\vec{B}_{ex}$ at $\alpha=4$.
In the absence of $\vec{B}_{ex}$, the total angular momentum $j_z=-\frac{1}{2}$
for small $\rho$ and decreases when $\rho \rightarrow 1$.
Introducing the field $\vec{B}_{ex}$ changes the ground state
dramatically.
If $\vec{B}_{ex}$ is parallel to the induced magnetic field $\vec{B}_R$, i.e.,
$\gamma>0$, $j_z$ first decreases then increases with the rotational
angular velocity $\rho$.
However, for $\gamma<0$, $j_z$ increases with $\rho$ monotonically.

\begin{figure}
\includegraphics[width=0.8\linewidth]{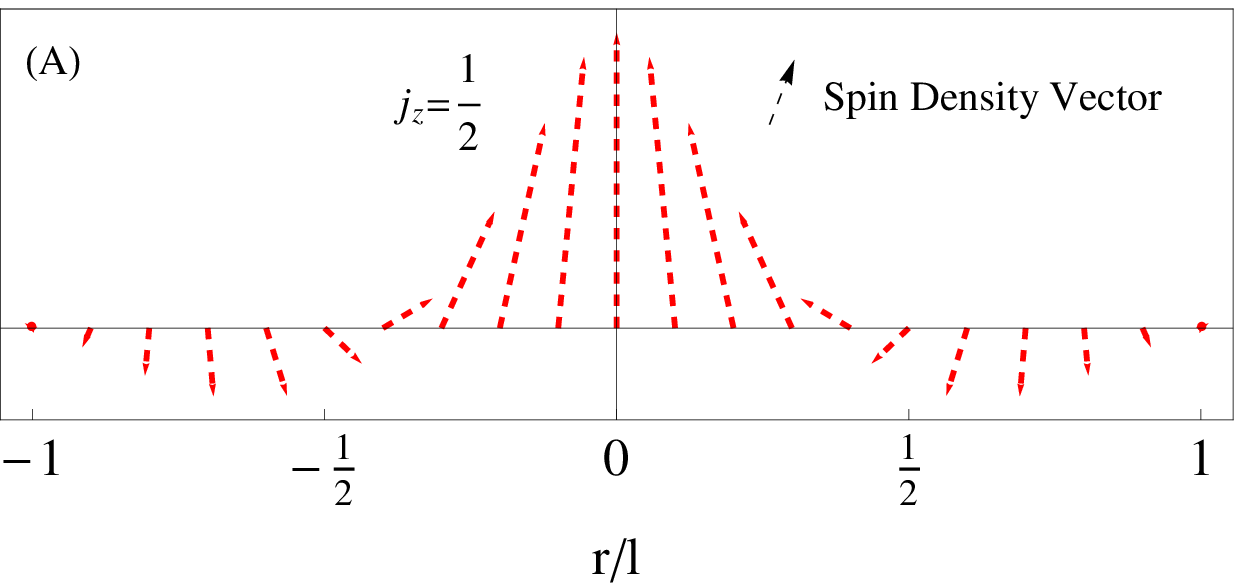} \\
\includegraphics[width=0.8\linewidth]{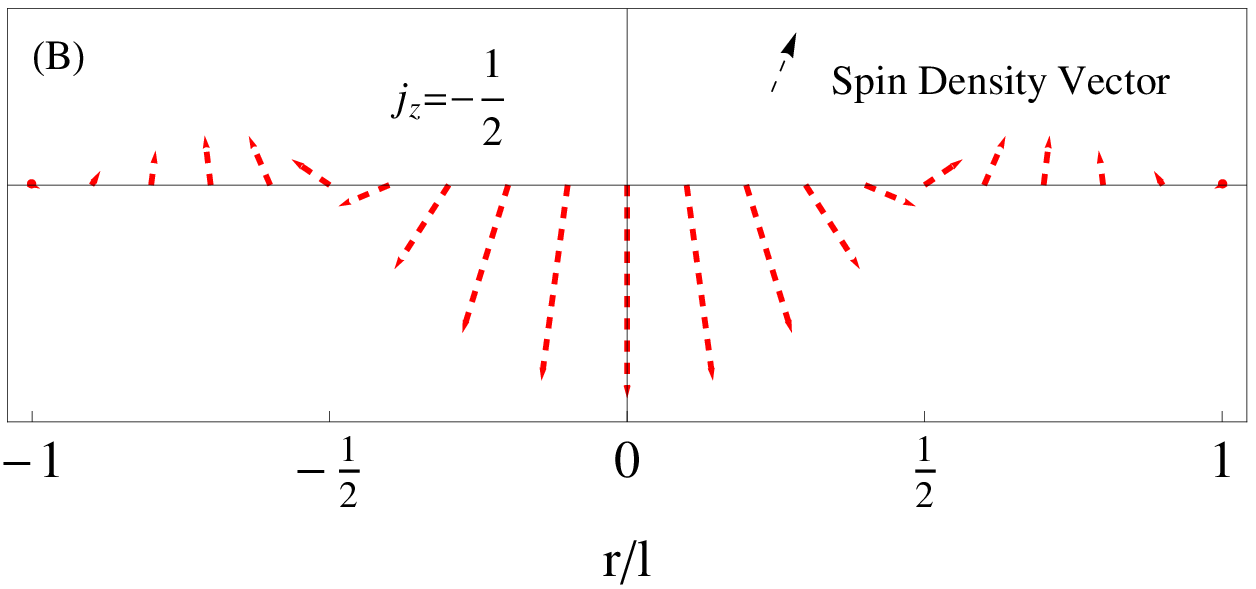}
\caption{The spin density vector distributions along the $x$-axis
lie in the $xz$-plane as shown in (A) $j_z=\frac{1}{2}$ and (B)
$j_z=-\frac{1}{2}$ with $\alpha=4$, and $\rho=\gamma=0$.
They are time-reversal counterparts to each other, and
both exhibit the skyrmion-type texture configuration.
}
\label{fig:spin_vec}
\end{figure}

The above results can be understood as follows.
In the case of $\Omega_z=0$, the two states $\phi_{j_z = \pm \frac{1}{2}}$ are
degenerated due to TR symmetry.
Since only one of the two spin components carries a vortex, the ground
state can be viewed as a half quantum vortex state with the density profiles
of two spin components shown in Fig. \ref{fig:single-particle-density}.
The spin density distributions exhibit skyrmion-type texture configurations,
as depicted in Fig. \ref{fig:spin_vec} A and B.
Intuitively, one might expect that an infinitesimal $\Omega_z$
selects the $\phi_{j_z = \frac{1}{2}}$ state
since it has lower rotational energy $-\Omega_z \langle L_z \rangle$.
However, the presence of the induced magnetic field $\vec{B}_R$
contributes another term to the total rotational energy of the
system as
\bea
\langle H_{rot} \rangle = -\Omega_z \langle L_z \rangle -
\vec{B}_R \cdot \langle \vec{\sigma} \rangle.
\eea
The spin pattern $\langle \vec{\sigma} \rangle$ for $\phi_{j_z = \frac{1}{2}}$ in the $xy$-plane is
anti-parallel to $\vec{B}_R$ near the trap center
(see Fig. \ref{fig:spin_vec}(A)), which is  energetically unfavorable.
Therefore, when $- \vec{B}_R \cdot \langle \vec{\sigma} \rangle$ dominates,
$j_z$ of the ground state can be $-\frac{1}{2}$ for a rotating trap.
As increasing $\Omega_z$, the condensates expand, which also favors
the magnetic energy term.
The total angular momentum $j_z$ can decrease even when $\rho$ increases.
Such a counter-intuitive effect for the ground state constitutes
a characteristic feature of SO coupled BECs in a rotating trap.
Introducing the external magnetic field $\vec{B}_{ex}$ strengthen
or weaken this effect induced by $\vec{B}_R$ depending  on its direction,
which explains the different behaviors of $m$ with $\rho$ for $\gamma>0$
and $\gamma<0$, as shown in Fig. \ref{fig:single_particle}.


\section{Vortex configurations of rotating SO coupled BEC}
\label{sect:vort}

Interaction effects in the absence of rotation have been investigated
extensively in the literature, which are summarized below.
In the case of a strong trapping potential and weak interaction,
the single-particle energy dominates.
The condensate maintains rotational symmetry but spontaneous breaks
TR symmetry \cite{wu2008,wu1}.
One spin-component carry one vortex, and the other is non-rotating,
thus the condensate possesses a half-quantum vortex.
The total angular momentum of each particle is $|j_z|=\frac{1}{2}$.
In momentum space, this kind of ground state distributes uniformly
around the Rashba ring.
On the contrary, if the trapping potential is weak and interaction
is strong, the condensate breaks rotational symmetry.
The condensate is approximately superposition of plane-wave states
modified by the cylindrical boundary condition.
Results based on the Gross-Pitaevskii (G-P) equation show that the
spin-spiral condensate with two counter-propagating plane-waves
is favored at $c>1$, while a single plane-wave is favored at $c<1$
\cite{ho,zhai,yip,zhang}.
These two different condensates are degenerate for the spin-independent
interactions, {\it i.e.}, $c=1$.
However, calculations including quantum fluctuations of the zero-point
energy show that the spin-spiral state wins at $c=1$, and thus shift the
phase boundary to a smaller value of $c$ \cite{wu1}.

In this section, we study the vortex configurations of SO coupled
BECs in both cases.
The results of strong trapping potentials and weak interactions
are presented in Sect. \ref{sect:trap}, and those of the opposite
limit are presented in  Sect. \ref{sect:weak_trap}.

\subsection{Vortex lattice configurations with a strong trapping potential}
\label{sect:trap}

\begin{figure}
\includegraphics[width=0.9\linewidth]{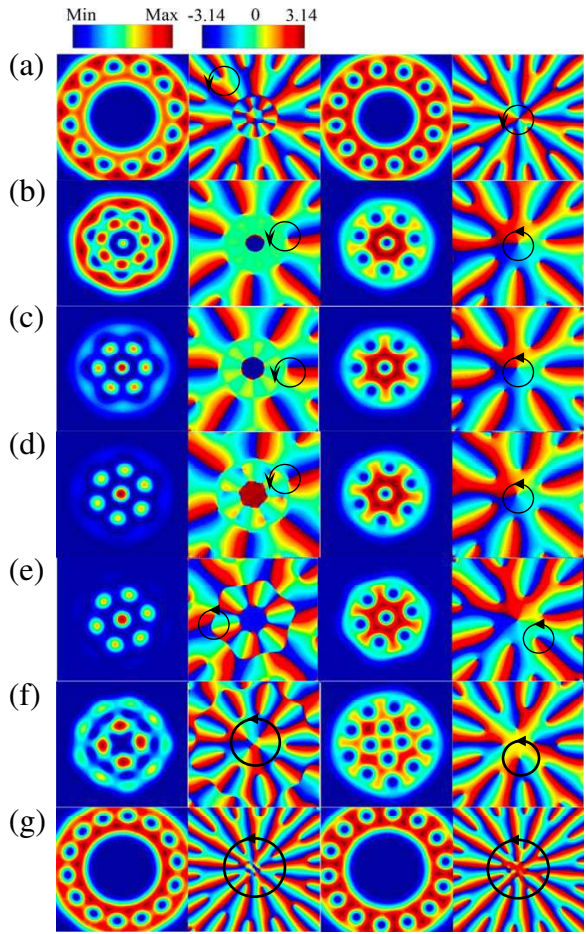}
\caption{
From left to right: the density and phase profiles of spin-up and down
components with parameter values of $\alpha=0.5$, $\beta=10$, $\rho=0.97$,
and $c=1$.
From (a)-(g), $\gamma$ is taken as $0.5$, $0.25$, $0.1$, $0.0$, $-0.1$, $-0.25$,
and $-0.5$, respectively.
At small values of $|\gamma|$ in (c) - (e), a half quantum vortex lattice
is formed near the trap center.
The spin-up component breaks into several density peaks, and the
low density region is connected.
As increasing the magnitude of $|\gamma|$ (b) and (f), the half-quantum vortex
lattice evolves to the normal vortex lattice.
For the large value of $|\gamma|=0.5$ (a) and (g),
the condensates show a lattice
configuration around
a ring. The black circle with an arrow indicates the the direction
of the circulation around the vortex core.
}
\label{fig:vort_latt}
\end{figure}

In this subsection, we turn on rotation and consider a strong
trapping potential with a small value of $\alpha$.
The ground state condensate is obtained by numerically solving the following
SO coupled GP equation.
We assume that the condensate is uniform along the $z$-axis, and
define the normalized condensate wavefunction
$(\tilde\psi_\uparrow, \tilde\psi_\downarrow)^T$
satisfying
$\int d^2 \vec r (|\psi_{\uparrow}|^2 + |\psi_{\downarrow}|^2)=1$.
The dimensionless version of the G-P equation can then be written as
\begin{subequations}
\label{eq:GP}
\bea
\frac{\mu}{\hbar \omega} \tilde{\psi}_{\uparrow}&=&\hat{T}_{\uparrow \nu}
\tilde{\psi}_{\nu} + \beta ( |\tilde{\psi}_{\uparrow}|^2 + c |\tilde{\psi}_{\downarrow}|^2 )
\tilde{\psi}_{\uparrow}, \\
\frac{\mu}{\hbar \omega} \tilde{\psi}_{\downarrow}&=&\hat{T}_{\downarrow \nu}
\tilde{\psi}_{\nu} + \beta ( |\tilde{\psi}_{\downarrow}|^2 + c |\tilde{\psi}_{\uparrow}|^2 )
\tilde{\psi}_{\downarrow},
\eea
\end{subequations}
where
\bea
\hat{T}&=&-\frac{1}{2}l^2(\partial^2_x + \partial^2_y) +
\alpha l (-i \partial_y \sigma_x + i \partial_x \sigma_y) \nn \\
&+& \frac{1}{2l^2}(x^2+y^2) - \rho (-i x\partial_y +iy\partial_x) \nn \\
&-& \frac{\alpha \kappa}{l} (x \sigma_x + y
\sigma_y),
\eea
where $\mu$ is the chemical potential; the interaction parameter
$\beta = g N /(\hbar \omega l_z)$; $N$ is the particle
number in the condensate; $l_z$ is the system size along the $z$-direction.

\begin{figure}
\includegraphics[width=0.9\linewidth]{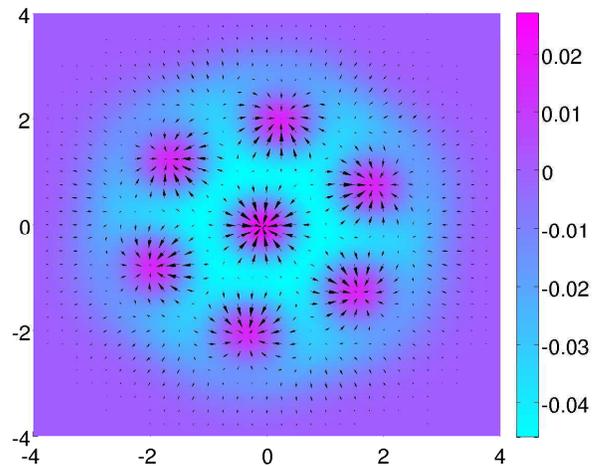}
\caption{
The ground-state spin density vector of Fig. \ref{fig:vort_latt} (d)
with parameter values of $\alpha=0.5$, $\beta=10$, $\rho=0.97$, $c=1$, and $\gamma=0$.
The projection of $\langle \vec{\sigma} \rangle$ in the $xy$-plane
is shown as black vectors. A color map is used to illustrate the $\langle \sigma_z \rangle$
component.
}
\label{fig:skymion}
\end{figure}

The density and phase configurations at various parameters are shown in
Fig. \ref{fig:vort_latt} (a) - (g), which exhibit rich structures of
vortex-lattice.
We look at Fig. \ref{fig:vort_latt} (d) in the absence of $\vec B_{ex}$,
{\it i.e.} $\gamma=0$.
The density distribution of the spin-up component is composed of several
disconnected density peaks near the trap center.
On the other hand, the low density region is connected in contrast
to the usual vortex lattice structure in which the low density
region of vortex cores is disconnected.
Nevertheless, we identify the locations of the singular points
of the phase distribution pattern around which the phase winds with
an integer number.
These singular points are squeezed out to the edge of the condensate.
On the other hand, the spin-down component exhibits the regular
vortex-lattice structure, whose vortex cores overlap with
the density peaks of the spin-up component.
Around each vortex core, the two spin components show a half-quantum vortex configuration as those depicted
in Fig. \ref{fig:single-particle-density}.
Therefore, the condensates of two components together
exhibit a lattice of half-quantum vortices.
The corresponding spin density vector $\langle \vec{\sigma} \rangle$ shows a skyrmion-lattice structure,
as shown in Fig. \ref{fig:skymion}.

Now we turn on the external Zeeman term Eq. \ref{eq:zeeman}.
For both cases of $\gamma>0$ and $\gamma<0$, at small values of
$|\gamma|$, the half-quantum vortex lattice still forms,
which is similar to that at $\gamma=0$
as depicted in Fig. \ref{fig:vort_latt} (b, c, e).
As increasing the strength of $\vec B_{ex}$, i.e., $|\gamma|$,
more vortices appear as depicted in Fig. \ref{fig:vort_latt} (b,f).
The condensates of the spin-up component gradually evolves to the
usual vortex-lattice configuration.
The high density region becomes connected, while the density minima
become disconnected vortex cores.
On the other hand, the condensates of the spin-down component remains
the usual vortex lattice configuration.
For even larger values of $|\gamma|$, the ring-shaped vortex lattice
with a giant vortex core is observed as shown in
Fig. \ref{fig:vort_latt} (a) and (g).
This is because the combined effect of the harmonic trap
$V_{ext}(\vec{r})$ and the additional Zeeman term $H_B$ shifts the
potential minimum to a ring in real space with the radius of $r=\alpha \gamma l=|B_0|/(M \omega^2)$.
The condensates of both spin up and down components distribute
around this ring and from a giant vortex configuration.
Additionally, the Zeeman term grows linearly
as increasing $r$ and favors in-plane polarization of $\vec{S}$.
As a result, the vortex cores of the spin up and down components
overlap with each other.

We stress that in all cases in Fig. \ref{fig:vort_latt} (a-g),
the vortex numbers in the spin-up and down components differ by one,
which is a characteristic feature brought by SO coupling.
As shown in Eq. \ref{eq:wavefunc}, for the eigenstate of the
single-particle Hamiltonian with $j_z=m+\frac{1}{2}$, the two
spin components carry different canonical orbital angular momenta
$m$ and $m+1$, respectively.
In the presence of interaction, the giant vortex splits into a
lattice of single-quantum vortices in each spin component.
Nevertheless, the total vortex number in each component remains
unchanged and differs by one.


\subsection{Weak trapping potential}
\label{sect:weak_trap}

\begin{figure}
\includegraphics[width=0.8\linewidth]{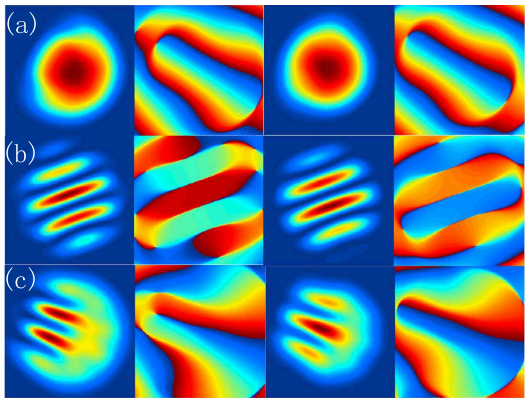}\\
\caption{From left to right: the density and phase profiles of spin up
and down components with the parameter values of $\alpha=4$, $\beta=20$,
and $\gamma=0$.
(a) $c=0.6$ and $\rho=0.1$, a plane-wave-like state is obtained with
a distorted phase pattern;
(b) $c=1.2$ and $\rho=0.1$, the spin-spiral condensate is favored;
(c) $c=1.2$ and $\rho=0.5$, the condensate exhibits an intermediate
configuration between those of (a) and (b).
The color scales for the density and phase distributions
are the same as those in Fig. \ref{fig:vort_latt}.
}
\label{fig:nonrot}
\end{figure}

\begin{figure}[!htbp]
\begin{center}
\includegraphics[width=0.95\linewidth]{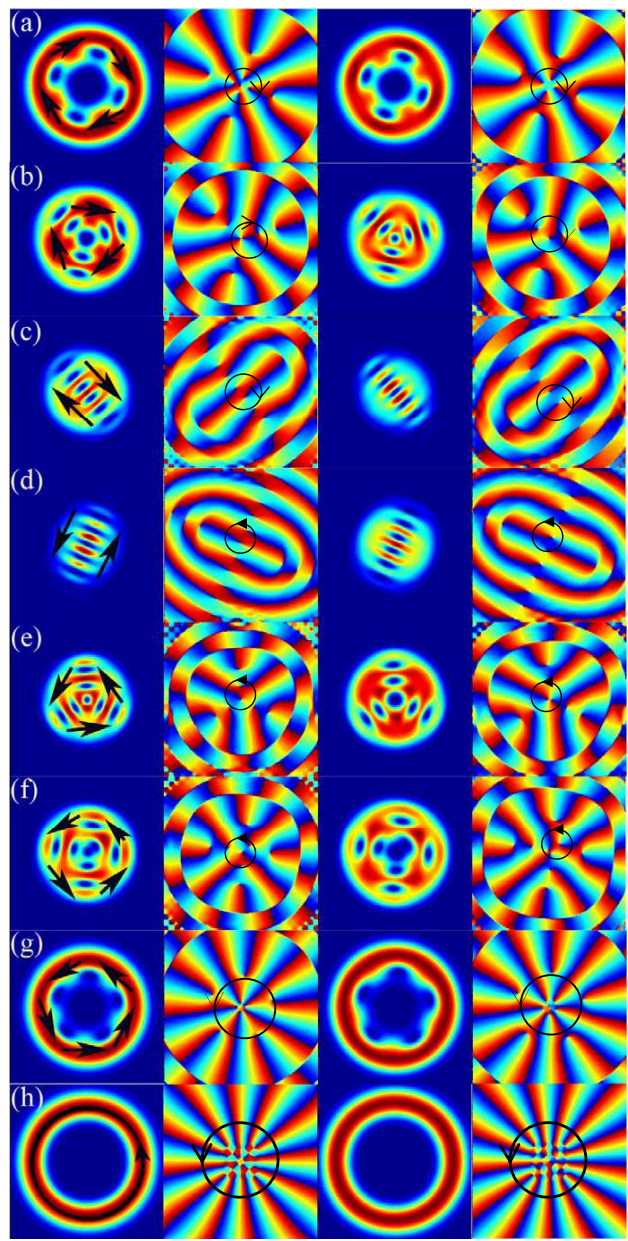}
\caption{From left to right: the density and phase profiles of spin up
and down components with parameter
values of $\alpha=4$, $\beta=20$, $c=1$, and $\rho=0.1$.
From (a)-(h), $\gamma$ is taken as $0.5$, $0.3$, $0.1$, $-0.05$,
$-0.25$, $-0.35$, $-0.6$, and $-0.7$, respectively.
The black arrow in each domain represents the local wavevector direction
of the corresponding plane-wave state, which shows a clockwise or
counter-clockwise configuration depending on the sign of $\gamma$.
For sufficiently large values of $|\gamma|$, condensates distribute
around a ring in space forming a giant vortex.
The color scales for the density and phase distributions are
the same as that in Fig. \ref{fig:vort_latt}. The black circle with an arrow indicates the the direction
of the circulation around the vortex core.}
\label{fig:domain}
\end{center}
\end{figure}

In this subsection, we study the rotating SO coupled BEC with a weak
trapping potential and strong interactions.

Fig. \ref{fig:nonrot} shows the density and phase profiles of each
spin component in the absence of external magnetic field $\vec{B}$,
i.e., $\gamma=0$.
In Fig. \ref{fig:nonrot} (a) with $c<1$, the condensate is a twisted
plane-wave state subject to the cylindrical boundary condition.
The spin polarization mainly lies in the $xy$-plane.
In the representation eigen-basis of $s_z$, the spin up and down
components show nearly the same distributions of density and phase profiles.
Nevertheless, the phase distribution is distorted from the
exact plane-wave state.
On the other hand, as depicted in Fig. \ref{fig:nonrot} (b), at $c>1$
the spin-spiral-like condensate with two counter-propagating plane-waves
is still favored with twisted phase profiles.
As shown in Fig. \ref{fig:nonrot} (c), increasing the angular velocity
$\rho$ gives rise to an intermediate configuration between
the distorted spin-spiral and the single-plane wave states.
In all the patterns, vortices locate either on the edge of the
condensate or the density minima of each component.


Next, we consider the case of $\gamma\neq 0$.
Introducing $H_B$ significantly enriches the structures of the
rotating SO coupled condensates.
We only consider a small angular velocity at $\rho=0.1$ for
the reason of numerical convergence, but vary the values of
$\gamma$ from $0.5\sim -0.7$ as presented in Fig. \ref{fig:domain}
(a)-(h), respectively.
With small and intermediate values of $|\gamma|$ (e.g. Fig. \ref{fig:domain}
(b)-(f)), the condensate breaks into several domains.
Inside each domain, the condensate can be approximated as a single
plane-wave state.
Vortices center around the local density minima.
The local wavevectors are configured such that the local spin
polarization $\avg{\vec S}$ align along the local Zeeman field
of $\vec{B}_{ex}(\vec r)$.
If $\gamma>0$ at which the external Zeeman field enhances the rotation
induced ones, we obtain a clockwise configuration of wavevectors.
There is one more vortex with the negative phase winding in the
spin up component than in the spin down component, which
reflects the ``anti-paramagnetic" feature.
On the contrary, if $\gamma<0$, the anti-clockwise patterns of wavevectors
is favored.
Similarly, the spin-down component also carries one more vortex
than the up component.

At small values of $|\gamma|$, two domains are formed as depicted
in Fig. \ref{fig:domain} (c) and (d).
The vortices   organize into straight-lines between two domains.
A variational wavefunction  is constructed as
\bea
\tilde{\psi}(\vec{r}) &\sim&  \left [  f_-(x) e^{-i \frac{\theta}{2}}
\psi_{-,-\vec{k}_0} +  f_+(x) e^{i \frac{\theta}{2}}
\psi_{-,\vec{k}_0} \right] \nn \\
&\times&
\frac{e^{-r^2/(2 a^2)}}{\sqrt{\pi}\sigma},
\eea
where without loss of generality, we choose the wavevector
$\vec{k}_0=k_0 \vec{e}_y$; $a$ is radius of the condensate;
$\theta$ is the relative phase difference between the two plane
wave domains; $|f_{-,+} (x)|^2 = (e^{\pm x/W} + 1)^{-1}$
are smeared step functions with $W$ the width of the domain wall.
We assume $\sigma \gg (W, 1/k_0)$. Such variational wavefunction has
neglectable contribution to the energy term $\avg{H_{rot}}$. This
explains why the two domain pattern is absent by increasing the rotational angular velocity $\rho$ only,
but appears immediately even at small values of $|\gamma|$.
With increasing $|\gamma|$, the condensate breaks into more and more
domains as in Fig. \ref{fig:domain} (b), (e) and (f).

As further increasing $|\gamma|$, domains connect together as a giant
vortex as shown in Fig. \ref{fig:domain} (a, g, h).
The condensates of both spin up and down components distribute
around a ring with the radius of $\alpha|\gamma|l$ and overlap each other.
This is a giant vortex configuration with a texture of spin
aligned along the radial direction.
The phase winding numbers of the spin-up and down components
differ by one due to the SO coupling.

\section{Conclusion}
\label{sect:conclusion}

To summarize, we have considered the vortex structures of SO coupled
BECs in a rotating trap combined with an external spatially dependent
Zeeman field.
In the case of strong confining potentials and weak interactions,
the condensate exhibit vortex-lattice structures.
As varying the  magnitude of the external Zeeman field,
the configuration evolves from a half-quantum vortex-lattice to
a normal one.
In the opposite limit, the condensate develops multi-domain patterns
with the external Zeeman field.
Each domain represents a local plane-wave state, whose wavevector
exhibit a clockwise or counter-clockwise configuration.
Domain boundaries play the role of like vortices.

\section*{Acknowledgement}
X. F. Z. gratefully acknowledges the support of NFRP (2011CB921204,
2011CBA00200), NNSF (60921091), NSFC (11004186), and CUSF, SRFDP
(20103402120031), and the China Postdoctoral Science Foundation.
C. W. is supported by NSF-DMR1105945 and AFOSR-YIP program.
C. W. acknowledge Aspen Center of Physics, where part of this
work was done there.
C. W. and X. F. Z. thank H. Pu, H. Hu, and X. J. Liu for helpful
discussions.

{\it Note added}~~
Near the completion of this manuscript, we notice a recent paper
studying the rotating Rashba SO coupled BEC, which considered
a special case in the presence of the extra term of Eq. \ref{eq:zeeman}
with $\gamma=-\rho$ \cite{xu2011}.
Our work has studied the general cases, including the pure
rotation without the external fields which corresponds
to $\gamma=0$.

\end{document}